\documentclass[aps,showpacs,nofootinbib,prd,twocolumn]{revtex4-1}
\usepackage{graphicx}
\usepackage{amssymb}
\usepackage{amsmath}
\usepackage{xcolor}
\usepackage{mathtools,slashed}
\usepackage{epstopdf}
\usepackage[utf8]{inputenc}
\usepackage{url}

\usepackage{newtxmath}

\usepackage{epsfig}

\newcommand{\be}{\begin{eqnarray}}
\newcommand{\ee}{\end{eqnarray}}
\newcommand{\NRECA}{N_{\bar{\Lambda} \,\mbox{\tiny{REC}}}}
\newcommand{\NREC}{N_{\Lambda \,\mbox{\tiny{REC}}}}
\newcommand{\NQGP}{N_{\Lambda \,\mbox{\tiny{QGP}}}}

\newcommand{\NpQGP}{N_{\text{p} \,\mbox{\tiny{QGP}}}}

\begin{document}

\title{Rise and fall of $\Lambda$ and $\overline{\Lambda}$ global polarization in semi-central heavy-ion collisions \\ at HADES, NICA and RHIC energies from the core-corona model}

\author{Alejandro Ayala$^{1,2}$}
\author{Isabel Dom\'inguez$^3$}
\author{Ivonne Maldonado$^{3,4}$}
\author{Mar\'ia Elena Tejeda-Yeomans$^5$}
  \address{
  $^1$Instituto de Ciencias
  Nucleares, Universidad Nacional Aut\'onoma de M\'exico, Apartado
  Postal 70-543, CdMx 04510,
  Mexico.\\
  $^2$Centre for Theoretical and Mathematical Physics, and Department of Physics,
  University of Cape Town, Rondebosch 7700, South Africa.\\
  $^3$Facultad de Ciencias F\'isico-Matem\'aticas, Universidad Aut\'onoma de Sinaloa,
Avenida de las Am\'ericas y Boulevard Universitarios, Ciudad Universitaria,
C.P. 80000, Culiac\'an, Sinaloa, Mexico.\\
  $^4$Joint Institute for Nuclear Research, Dubna, 141980 Russia.\\
  $^5$Facultad de Ciencias - CUICBAS, Universidad de Colima, Bernal D\'iaz del Castillo No. 340, Col. Villas San Sebasti\'an, 28045 Colima, Mexico.
  }
\begin{abstract}
    We compute the $\Lambda$ and $\overline{\Lambda}$ global polarizations in semi-central heavy-ion collisions using the core-corona model where the source of $\Lambda$'s and $\overline{\Lambda}$'s is taken as consisting of a high-density core and a less dense corona. We show that the overall properties of the polarization excitation functions can be linked to the relative abundance of $\Lambda$s coming from the core versus those coming from the corona. For low collision energies, the former are more abundant whereas for higher energies the latter become more abundant. The main consequence of this reversing of the relative abundance is that both polarizations peak at collision energies $\sqrt{s_{NN}}\lesssim 10$ GeV. The exact positions and heights of these peaks depend not only on this reversal of relative abundances, but also on the centrality class, which is directly related to the QGP volume and lifetime, as well as on the relative abundances of $\Lambda$s and $\overline{\Lambda}$s in the core and corona regions.
    The intrinsic polarizations are computed from a field theoretical approach that links the alignment of the strange quark spin with the thermal vorticity and modeling the QGP volume and lifetime using a Bjorken expansion scenario. We predict that the $\Lambda$ and $\overline{\Lambda}$ global polarizations should peak at the energy range accessible to NICA and HADES.
\end{abstract}
\maketitle
\section{Introduction}

The polarization properties of $\Lambda$ and $\overline{\Lambda}$ have received increasing attention over the last years due to the possibility to
link this observable to the properties of the medium produced in relativistic heavy-ion collisions~\cite{Jacob,Barros,Ladygin,Becattini1,Xie,Karpenko:2016jyx,Xie2,Liao,Liao2,Li,Karpenko2,Xia,Suvarieva}. For semi-central collisions, the matter density profile in the transverse plane develops an angular momentum~\cite{Becattini2008} which can be quantified in terms of the thermal vorticity~\cite{Becattini:2016gvu}. When this vorticity is transferred to spin degrees of freedom, the global polarization can be measured using the self-analising $\Lambda$ and $\overline{\Lambda}$ decays. A significant effort has been devoted to study both the local and global polarization of these hyperons that could be produced by this vorticity in heavy-ion reactions~\cite{Karpenko:2013wva,DelZanna:2013eua,Karpenko:2016jyx,Becattini:2016gvu,Ivanov:2019ern,Wei:2018zfb,Vitiuk:2019rfv,Xie:2019jun,Ivanov:2020udj,Karpenko:2021wdm}. In particular, hydrodynamical simulations, that successfully describe flow observables and hadron abundances at RHIC energies, have been put to the test in an effort to understand the rise of $\Lambda$ and $\overline{\Lambda}$ polarization at lower collision energies~\cite{STAR:2021beb}.

The Beam Energy Scan (BES) at RHIC, performed by the STAR Collaboration~\cite{STAR-Nature,STAR2,STAR:2021beb} has shown a trend for the $\Lambda$ and $\overline{\Lambda}$ global polarization to increase as the energy of the collision decreases and that this increase is faster for $\overline{\Lambda}$s than for $\Lambda$s. In addition, the HADES Collaboration has recently provided preliminary results on the $\Lambda$ global polarization in Au+Au collisions at $\sqrt{s_{NN}}= 2.42$ GeV~\cite{SQM2019} finding a non-vanishing result.

The theoretical and phenomenological ideas to explain the properties of hyperon global polarization follow different and partially successful avenues. The models and simulations providing hyperon polarization predictions depend on control parameters such as the colliding energy and beam species but more importantly, on the main polarization  driving mechanism. The STAR-BES results seem to indicate that this mechanism needs to differentiate between hyperons and anti-hyperons.

Among the mechanisms to explain the difference in the global $\Lambda$ and $\overline{\Lambda}$ polarization one can mention possible different space-time distributions and freeze-out conditions for $\Lambda$ and $\overline{\Lambda}$~\cite{Vitiuk:2019rfv}; the polarization of $s$ and $\bar{s}$-quarks induced by short-lived but intense magnetic fields~\cite{Hai-Bo,Liao3,Guo:2021uqc,Liao4}; the possibility that $\Lambda$ and $\overline{\Lambda}$ align their spins with the direction of the angular momentum created in the reaction during the life-time of the evolving system~\cite{Ayala,Ayala:2020ndx} and
a dynamical mechanism with an interaction, mediated by massive vector and scalar bosons, between the spins of hyperons and antihyperons and the vorticity of the baryon current~\cite{Csernai:2018yok,Xie:2019wxz}.

In a recent work~\cite{Ayala:2020soy}, we expanded on the idea, first put forward in Ref.~\cite{Ayala:2001jp} and later on also studied in Refs.~\cite{Werner:2007bf, Aichelin:2008mi}, that in semi-central collisions, $\Lambda$s and $\bar{\Lambda}$s can be produced in different density zones within the reaction volume. A similar idea was also discussed in Ref.~\cite{Baznat:2015eca}. We have shown that by modeling the source of $\Lambda$s and $\overline{\Lambda}$s as consisting of a high-density core and a less dense corona, the global polarization properties of these hyperons, as functions of the collision energy, are well described. The quark gluon plasma (QGP) is produced in the core only when the density of participants in the colliding nuclei exceeds a critical value. On the other hand, in the corona, the density of participants is smaller than this critical value and particle production processes are similar to those in $p + p$ reactions. For a given impact parameter (or rather, a centrality class), the volume in the corona becomes larger at lower energies. We found that when the larger abundance of $\Lambda$s compared with $\overline{\Lambda}$s coming from the corona is combined with a smaller number of $\Lambda$s coming from the core, compared with those from the corona, which happens for collisions with intermediate to large impact parameters, an amplification effect for the $\overline{\Lambda}$ polarization can occur, in spite of the  intrinsic $\Lambda$ polarization $z$ being larger than the intrinsic $\overline{\Lambda}$ polarization $\bar{z}$. This amplification is more prominent for lower collision energies. The model provided a good description of the different increasing trends of $\Lambda$/$\overline{\Lambda}$ polarization measured by the STAR-BES at RHIC. The purpose of this work is to use and improve the model to predict the polarization of these hyperons for NICA and HADES energies. As we show, the model predicts that both polarizations peak in this energy region to then decrease and become zero near the threshold  energy for $\Lambda$/$\overline{\Lambda}$ production. This result is in agreement with the recent preliminary results reported by HADES for Au+Au collisions at $\sqrt{s_{NN}}= 2.42$ GeV~\cite{SQM2019} and by the STAR-BES at $\sqrt{s_{NN}}=3$ GeV~\cite{STAR:2021beb}.

We notice that the existence of a peak in the polarization excitation functions has also been found using hydrodynamical and transport calculations, extrapolated to low energies. These calculations include  the three-Fluid Dynamics (3FD) model~\cite{Ivanov:2020udj}, UrQMD~\cite{Deng:2020ygd} and AMPT~\cite{Guo:2021uqc}. However, only the 3FD model agrees well with data over the  analyzed energy range, although it overshoots the reported polarization value for $\sqrt{s_{NN}} = 3$ GeV~\cite{STAR:2021beb}. Using this model, the position of the peak of the $\Lambda$ polarization function is located at the same energy that what we find in this work. However, the $\overline{\Lambda}$ polarization trend is not reproduced.

The work is organized as follows: In Sec.~\ref{II} we describe the improved core-corona model and show how knowledge of the relative $\Lambda$ abundances in one and the other regions makes it possible to understand the rise and fall of the global polarization as a function of the collision energy. In Sec.~\ref{III} we compute the intrinsic polarization from a field theoretical calculation of the rate for the spin alignment with the thermal vorticity and from a simple space-time picture for the volume and life-time of the QGP evolution with collision energy. Putting all the ingredients together, the results are shown and discussed in Sec.~\ref{IV}. We finally summarize and conclude in Sec.~\ref{V}.

\section{Improved core-corona model}\label{II}

\begin{figure}[t]
    \centering
    \includegraphics[width=0.48\textwidth]{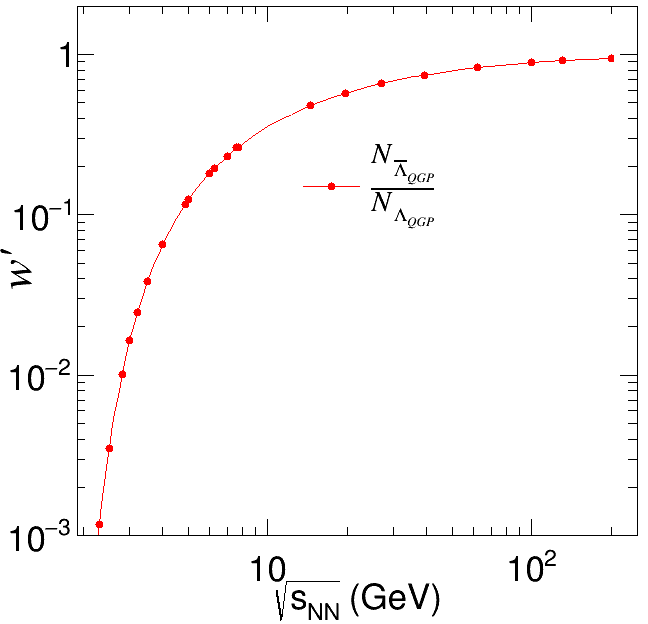}
    \caption{The ratio $w'=N_{\overline{\Lambda}\ {\mbox{\tiny{QGP}}}}/N_{\Lambda\ {\mbox{\tiny{QGP}}}}$ given by Eqs.~(\ref{equilibriumratio}) and~(\ref{Tmurel}) as a function of $\sqrt{s_{NN}}$.}
    \label{fig:ratio}
\end{figure}

The core-corona model, developed in Ref.~\cite{Ayala:2020soy}, provides a framework to compute the $\Lambda$ and $\overline{\Lambda}$ polarizations as
\begin{eqnarray}
\mathcal{P}^\Lambda=\frac{z\frac{
N_{\Lambda\ {\mbox{\tiny{QGP}}}} }{N_{\Lambda\ {\mbox{\tiny{REC}}}}}}{ \left( 1 + \frac{N_{\Lambda\ {\mbox{\tiny{QGP}}}}}{N_{\Lambda\ {\mbox{\tiny{REC}}}}}\right)},\ \ \
\mathcal{P}^{\overline{\Lambda}}=\frac{\left(\frac{\bar{z}}{w}\right)
\frac{N_{\Lambda\ {\mbox{\tiny{QGP}}}} }{N_{\Lambda\ {\mbox{\tiny{REC}}}}}}{ \left( 1 + 
\left(\frac{1}{w}\right)
\frac{N_{\Lambda\ {\mbox{\tiny{QGP}}}}}{N_{\Lambda\ {\mbox{\tiny{REC}}}}}\right)},
\label{eq1}
\end{eqnarray}
which depend on the number of $\Lambda$s produced in the core $\NQGP$, and in the corona $\NREC$. The subscripts \lq\lq QGP" and \lq\lq REC" refer to the kind of processes that mainly take place for the production of these hyperons; coalescence-type of processes in the QGP and recombination of a di-quark (antiquark) with an s-quark (antiquark). The notation is the one used to describe these processes in Ref.~\cite{Ayala:2001jp}. $w$ is the ratio between the number of $\bar{\Lambda}$s and $\Lambda$s created in the corona region, namely $w=\NRECA/\NREC$, 
and $z$ and $\bar{z}$ are the intrinsic $\Lambda$ and $\bar \Lambda$ polarization,  respectively, which are produced in the core, given that in the corona cold nuclear matter reactions are less efficient to produce an alignment between the $s$-quark (antiquark) spin and the thermal vorticity. 

One of the assumptions leading to Eqs.~(\ref{eq1}) is that in the core, QGP-like processes make it equally as easy to produce $\Lambda$s and $\overline{\Lambda}$s, given that in this region quarks and antiquarks are freely available and three antiquarks ($\bar{u}$, $\bar{d}$, $\bar{s}$) can find each other as easily as three quarks ($u,\ d,\ s$). To improve the model, we first notice that to account for a possible bias in the production of $\Lambda$s versus $\overline{\Lambda}$s, introduced by a more abundant production of $s$ over $\bar{s}$ at a finite value of the chemical potential, we can relax this assumption by writing \begin{eqnarray}
N_{\overline{\Lambda}\ {\mbox{\tiny{QGP}}}}=w'N_{\Lambda\ {\mbox{\tiny{QGP}}}}.
\label{relax}
\end{eqnarray}
The factor $w'$ is computed as the ratio of the equilibrium distributions of $\bar{s}$ to $s$ for a given temperature and chemical potential $\mu=\mu_B/3$, namely
\begin{eqnarray}
w'=\frac{e^{(m_s-\mu)/T}+1}{e^{(m_s+\mu)/T}+1},
\label{equilibriumratio}
\end{eqnarray}
where $m_s=100$ MeV is the $s$-quark mass, $T$ and $\mu_B$ (given in MeV) are taken as the values along the maximum chemical potential curve at freeze-out by~\cite{Cleymans}
\begin{eqnarray}
T(\mu_B)&=&166 - 139\mu_B^2 - 53\mu_B^4,\nonumber\\
\mu_B(\sqrt{s_{NN}})&=&\frac{1308}{1000+0.273\sqrt{s_{NN}}},
\label{Tmurel}
\end{eqnarray}
as a function of $\sqrt{s_{NN}}$. The ratio $w'$ is shown in Fig.~\ref{fig:ratio} as a function of $\sqrt{s_{NN}}$. Notice that $w'$ quickly drops down to zero in the NICA/HADES energy ranges. Using Eq.~(\ref{relax}) into Eq.~(\ref{eq1}), the polarization expressions are given now as
\begin{eqnarray}
\mathcal{P}^\Lambda=\frac{z\frac{
N_{\Lambda\ {\mbox{\tiny{QGP}}}} }{N_{\Lambda\ {\mbox{\tiny{REC}}}}}}{ \left( 1 + \frac{N_{\Lambda\ {\mbox{\tiny{QGP}}}}}{N_{\Lambda\ {\mbox{\tiny{REC}}}}}\right)},\,\,\,
\mathcal{P}^{\overline{\Lambda}}=\frac{\bar{z}\left(\frac{w'}{w}\right)
\frac{N_{\Lambda\ {\mbox{\tiny{QGP}}}} }{N_{\Lambda\ {\mbox{\tiny{REC}}}}}}{ \left( 1 + 
\left(\frac{w'}{w}\right)
\frac{N_{\Lambda\ {\mbox{\tiny{QGP}}}}}{N_{\Lambda\ {\mbox{\tiny{REC}}}}}\right)}.
\label{eq1-mod}
\end{eqnarray}
\begin{figure}[t!]
    \centering
    \includegraphics[width=0.48\textwidth]{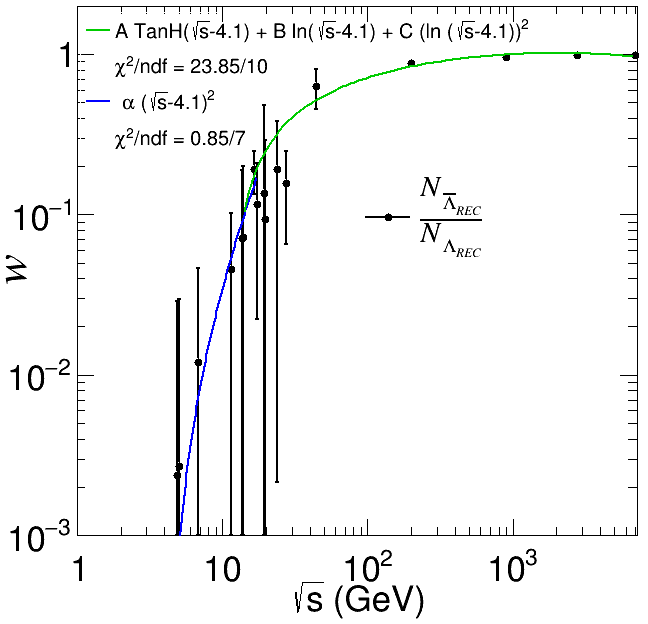}
    \caption{Experimental data obtained from p + p collisions at different energies~\cite{Gazdzicki:1996pk,Blobel:1973jc,Chapman:1973fn,Brick:1980vj,Hohne:2003bca,Baechler:1991pp,Charlton:1973kw,Lopinto:1980ct,Kichimi:1979te,Busser:1975tj,Erhan:1979ba,Abelev:2006cs,Abbas:2013rua}, fitted with the function  $w = \alpha \left(   \sqrt{s} - 4.1 \right)^2$ for $\sqrt{s}< 17.3$  GeV (blue line) and $w = A \tanh{(\sqrt{s}-4.1)} + B \ln{(\sqrt{s}-4.1)} + C \ln^2{(\sqrt{s}-4.1)}$ for $\sqrt{s} > 14$ GeV (green line). See the text for the values of the fit parameters.}
    \label{fig:wfit}
\end{figure} 

Notice that in the corona, $\Lambda$ and $\overline{\Lambda}$ producing reactions are similar to those in $p + p$ collisions, where it is easier to produce $\Lambda$s than $\overline{\Lambda}$s. Therefore, $w$ can be obtained from experimental data on $p + p$ collisions as a function of the center of mass energy $\sqrt{s}$ and it is expected to be less than 1. Figure~\ref{fig:wfit} shows a compilation of the $\Lambda/\overline{\Lambda}$ ratio in $p + p$ reactions in the energy range 4.86 GeV $< \sqrt{s}<$ 7 TeV~\cite{Gazdzicki:1996pk,Blobel:1973jc,Chapman:1973fn,Brick:1980vj,Hohne:2003bca,Baechler:1991pp,Charlton:1973kw,Lopinto:1980ct,Kichimi:1979te,Busser:1975tj,Erhan:1979ba,Abelev:2006cs,Abbas:2013rua}. Shown are also separate fits to the experimental ratio. The fits assume that $w$ is defined only for $\sqrt{s} > 4.1$ GeV which is the threshold energy to produce a $\bar{\Lambda}$ by means of the reaction $p + p \rightarrow p + p + \Lambda\ +\ \bar{\Lambda}$. For low energies (blue line) $\sqrt{s} < 15$ GeV, the data are fit with the function
$w = \alpha \left(   \sqrt{s} - 4.1 \right)^2$, where $\alpha = 0.0010 \pm 0.0003$. For higher energies (green line) $\sqrt{s} > 15 $ GeV, the data are fit with the function $w = A \tanh{(\sqrt{s}-4.1)} + B \ln{(\sqrt{s}-4.1)} + C \ln^2{(\sqrt{s}-4.1)}$, where $A = -0.8603 \pm 0.0965$, $B = 0.4935 \pm 0.0314$ and $C = -0.0324 \pm 0.0024$. 

Notice that the experimental results support the expectation that $w<1$. 

To estimate the number of $\Lambda$s produced in the core and the corona, we introduce a critical density of participants $n_c=3.3$ fm$^{-2}$ above (below) which, the QGP is (is not) formed. Then the number of $\Lambda$s from the core, $\NQGP$, is proportional to the number of participant nucleons in the collision above this critical value, $\NpQGP$, which is given by
\begin{eqnarray}
   \NpQGP = \int d^2s\ n_\text{p}(\Vec{s},\Vec{b})\,\theta \left[n_\text{p}(\Vec{s},\Vec{b})-n_c\right],
\label{numberof participants}    
\end{eqnarray}
where the density of participants $n_\text{p}$ is given in terms of the thickness functions $T_A$ and $T_B$ of the colliding system $A+B$ as
\begin{eqnarray}
   n_\text{p}(\Vec{s},\Vec{b})&=&T_A(\vec{s}\,)[1-e^{-\sigma_{NN}(\sqrt{s_{NN}})T_B(\vec{s}-\vec{b})}]\nonumber\\
   &+&T_B(\vec{s}-\vec{b})[1-e^{-\sigma_{NN}(\sqrt{s_{NN}})T_A(\vec{s})}],
\label{np}
\end{eqnarray}
with $\vec{b}$ the vector directed along the impact parameter on the nuclei overlap area and $\sigma_{NN}$ the collision energy-dependent nucleon + nucleon (N + N) cross-section. The thickness function $T_A$ is given by
\begin{eqnarray}
   T_A(\vec{s}\,)=\int_{-\infty}^{\infty}\rho_A(z,\vec{s}\,)\;dz,
\label{TA}
\end{eqnarray}
where we take as the nuclear density $\rho_A$ a Woods-Saxon profile with a skin depth $a = 0.523$ fm and a radius $R = 6.554$ fm \cite{Adamczewski-Musch:2017sdk, Kardan:2015}. With this information at hand, we  can estimate the average number of strange quarks produced in the QGP, and thus the number of $\Lambda$s, as a quantity that scales with the number of participants $\NpQGP$ in the collision, as

\begin{eqnarray}
\langle s\rangle = \NQGP = c\, \NpQGP^2,
\label{NLQGP}
\end{eqnarray} where we use $c=0.0025$~\cite{Ayala:2020soy}.

\begin{figure}[t]
    \centering
    \includegraphics[width=0.45\textwidth]{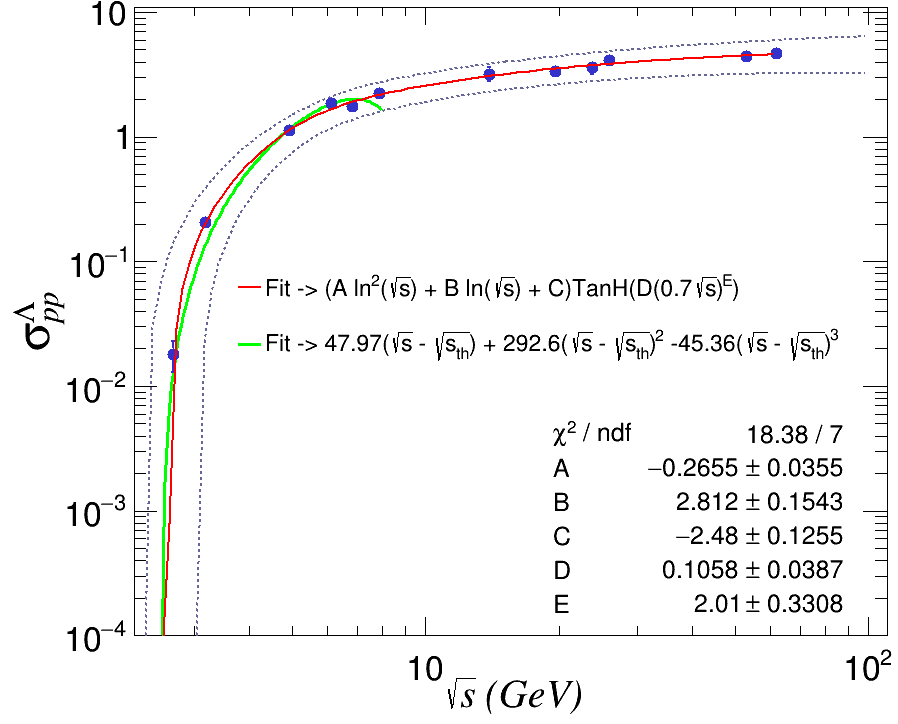}
    \caption{Fit (red) to the hyperon production cross-section in p + p collisions as a function of $\sqrt{s}$ using data reported in Refs.~\cite{Gazdzicki:1996pk, Blobel:1973jc, Chapman:1973fn, Brick:1980vj, Erhan:1979ba, Fickinger:1962zz, AdamczewskiMusch:2016vrc, Aahlin:1977vh, Boeggild:1973ex, Bogolyubsky:1988ei, Jaeger:1974in, Sheng:1974zn, Asai:1984dv,Drijard:1981wq}. Fit (green) to the hyperon production cross section in p + p collisions for near threshold energies reported by the HADES Collaboration~\cite{AdamczewskiMusch:2021rlv}.}
    \label{fig:sigmafit}
\end{figure}

Now, to compute the number of $\Lambda$s produced in the corona, $\NREC$, we note that the $\Lambda/\overline{\Lambda}$ production mechanism is the same as in N + N collisions, when the density of participants in the collision region is less than the critical density $n_p$.

Therefore, we can write
the number of $\Lambda$s produced in the corona as
\begin{eqnarray}
   \NREC&=& \sigma_{NN}^\Lambda\left(\sqrt{s_{NN}}\right)
   \int d^2s\; T_B(\vec{b}-\vec{s}) \nonumber \\
   &\times& T_A(\vec{s}\,)\,\theta \left[n_c-n_\text{p}(\Vec{s},\Vec{b})\right],
\label{partper}
\end{eqnarray}
For the $N + N$ cross-section for $\Lambda$ production we use the $p + p$ cross section $\sigma_{pp}^{\Lambda}$, which is a collision energy dependent quantity that can be obtained from a fit to data. In Fig.~\ref{fig:sigmafit}, we show a compilation of experimental data for $\sigma_{pp}^{\Lambda}$, covering a wide range of energies from a few to almost 70 GeV~\cite{Gazdzicki:1996pk, Blobel:1973jc, Chapman:1973fn, Brick:1980vj, Erhan:1979ba, Fickinger:1962zz, AdamczewskiMusch:2016vrc, Aahlin:1977vh, Boeggild:1973ex, Bogolyubsky:1988ei, Jaeger:1974in, Sheng:1974zn, Asai:1984dv, Drijard:1981wq}. A fit to these data is also shown in  Fig.~\ref{fig:sigmafit} with the red continuous curve inside the band, whose width represents the fit uncertainty. Notice that for the HADES collision energy, $\sqrt{s}=2.42$ GeV, the fit yields a negative value for the cross section at an energy just below the $\Lambda$ production threshold energy $\sqrt{s_{\mbox{th}}}\approx 2.55$ GeV for the reaction $p + p$ $\rightarrow$ K$^+ + \Lambda + p$. Thus, for energies below $\sqrt{s_{\mbox{th}}}$, we take this cross section as being zero. The vanishing of the cross section means that near threshold the produced  $\Lambda$s come mainly from the core region. The cross-section for the $p + p$ $\rightarrow$ K$^+$ + $\Lambda$ + $p$ exclusive channel has been measured at energies $\sqrt{s} = 2.549,\ 2.602,\ 2.805$ GeV by the COSY Collaboration~\cite{Balewski:1991ns,Bilger:1998jf} A recent fit of the hyperon production cross section in $p + p$ collisions for near threshold energies has been provided by the HADES Collaboration~\cite{AdamczewskiMusch:2021rlv}. This fit is shown by the green continuous line in Fig.~\ref{fig:sigmafit}. Notice that  in the restricted energy range from threshold to about 10 GeV, both fits are consistent with each other.

\begin{figure}
    \centering
    \includegraphics[width=0.45\textwidth]{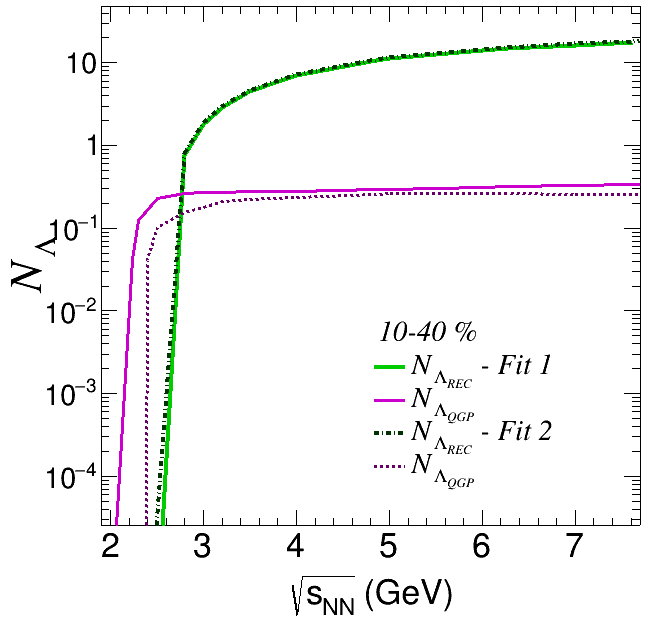}
    \caption{Number of $\Lambda$s using two different parametrizations of $\sigma_{NN}$ computed for $\langle b \rangle = 7.26$ fm, corresponding to the average impact parameter in the 10\%-40\%  centrality class. Notice that for $\Lambda$s produced in the QGP, both fits give similar results for $\sqrt{s_{NN}} > 3$ although for lower energies they differ.}
    \label{fig:NumberHyperons2}
\end{figure}

Finally, to evaluate the number of $\Lambda$s both in the core and the corona region, we also need the $\sigma_{NN}$ collision energy-dependent $N + N$ cross-section, that appears in Eq.~(\ref{np}). For $\sqrt{s} > 5$ GeV we can use the standard PDG parametrization~\cite{Nakamura:2010zzi}. However this parametrization is not suited for low energies, therefore the need to employ a different parametrization. Given that the experimental information on this cross-section is scarce, here we present results based on two different fits. The first one (\textit{Fit 1}) 
is taken from reference~\cite{Buss:2011mx} and the second one (\textit{Fit 2}) from Ref.~\cite{Bystricky:1987yq}. The resulting number of $\Lambda$s/$\overline{\Lambda}$s is shown in Fig.~\ref{fig:NumberHyperons2}. Notice that for $\sqrt{s_{NN}} > 3$ GeV the obtained number of $\Lambda$s in the corona is similar for both fits. However the number of $\Lambda$'s in the QGP is smaller for the second fit and goes to zero at $\sqrt{s_{NN}} \simeq 2.3$ GeV whereas for the first fit it vanishes at $\sqrt{s_{NN}} \simeq 2.1$ GeV. This difference impacts our determination of the $\Lambda$/$\overline{\Lambda}$ polarization strength and, correspondingly, we will show our results using both fits.

\begin{figure}[t]
    \centering
    \includegraphics[width=0.45\textwidth]{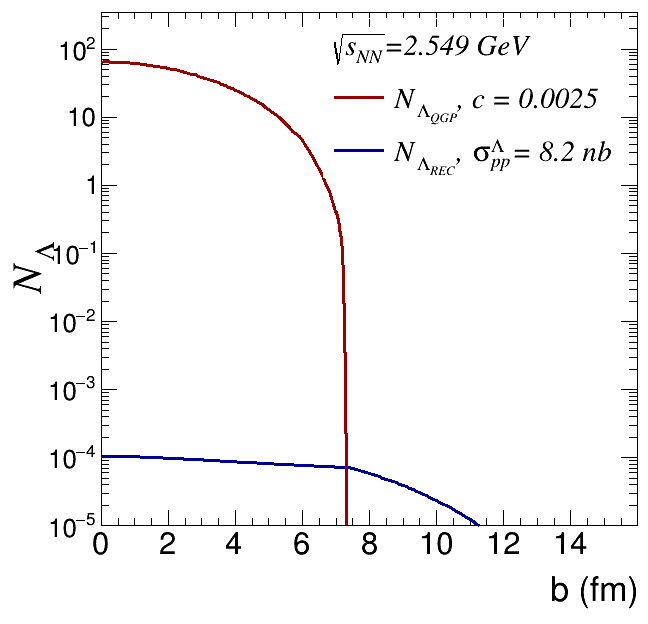}
	\caption{Number of $\Lambda$s created in the core and the corona, with $\sigma^{\Lambda}_{pp} = 8.2$\ nb and $\sigma_{NN}= 23.8$ mb. The region of interest $\left< b \right> = 7.26$ fm is where distributions are similar.} 
    \label{fig:NumberHyperons}
\end{figure}

\begin{figure}[b]
    \centering
    \includegraphics[width=0.45\textwidth]{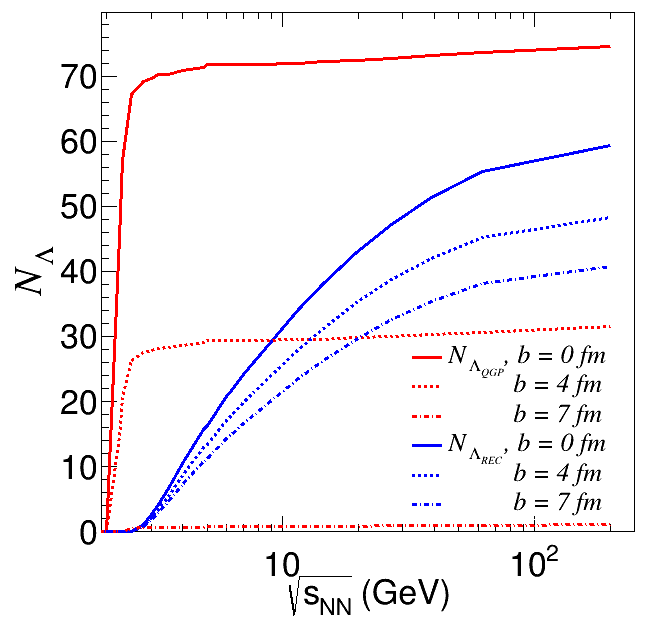}
	\caption{Number of $\Lambda$s created in the corona $\NREC$ (blue) and the core $\NQGP$ (red), as a function of the collision energy, for fixed impact parameters $b=0,4,7$ fm, that represent different centralities.} 
    \label{fig:NumberHyperons3}
\end{figure}

As an example, Fig.~\ref{fig:NumberHyperons} shows the number of $\Lambda$s created in the two regions as a function of the impact parameter for a collision energy with $\sqrt{s_{NN}}=2.549$ GeV. We have taken $\sigma^{\Lambda}_{pp}$ as the lowest measured  value by the COSY-TOF experiment. We observe that any change in the value of $\sigma^{\Lambda}_{pp}$ affects the ratio $\NQGP/\NREC$ and the value of the impact parameter $b$ at which the ratio is smaller than 1. In Fig.~\ref{fig:NumberHyperons3} we show the number of $\Lambda$s created in the core and the corona, as a function of the collision energy, for fixed impact parameters $b=0,4,7$ fm, that in turn correspond to different centralities. Notice that, whereas at small impact parameters, particle production is dominated by the core region, for peripheral collisions, relevant for vorticity and polarization studies, the situation reverses, and particle production becomes dominated by the corona region. It is easy to understand the origin of this behavior: core-corona models introduce a critical density of participants ($n_c$) above which the core can be produced. For peripheral collisions this critical density is difficult to be achieved, even for the largest collision energies.

From Eq.~(\ref{eq1-mod}), we notice that knowledge of the $\Lambda$ abundances in the core and the corona as functions of the control parameters, allows us to estimate the general behavior of the ratios of global to intrinsic polarizations ${\mathcal{P}}/z$ and $\overline{{\mathcal{P}}}/\bar{z}$ as functions of collision energy. 
These functions are controlled by the product of the monotonically decreasing ratios $N_{\Lambda\ {\mbox{\tiny{QGP}}}}/N_{\Lambda\ {\mbox{\tiny{REC}}}}$, $N_{\overline{\Lambda}\ {\mbox{\tiny{QGP}}}}/N_{\overline{\Lambda}\ {\mbox{\tiny{REC}}}}$ and the monotonically increasing ratios $1/(1+N_{\Lambda\ {\mbox{\tiny{QGP}}}}/N_{\Lambda\ {\mbox{\tiny{REC}}}})$, $1/(1+N_{\overline{\Lambda}\ {\mbox{\tiny{QGP}}}}/N_{\overline{\Lambda}\ {\mbox{\tiny{REC}}}})$, respectively. These products start growing from the lowest collision energy considered in this work, namely the one corresponding to the Lambda production threshold $\sqrt{s_{NN}}=2.54$ GeV up to an energy $\sqrt{s_{NN}}\simeq 2.8,\ 6.7$ GeV, respectively, where they reach a maximum to then start decreasing and become of order $10^{-2}$ already for RHIC energies. When these ratios are multiplied by $z$ or $\bar{z}$, respectively, the position of the corresponding peak is slightly displaced, as these latter factors have a mild energy dependence. To have an accurate estimate of the peaks position and shape of the polarization functions, we now proceed to describe the calculation of the intrinsic polarizations $z$ and $\bar{z}$.

\section{Intrinsic polarizations from spin alignment with vorticity}\label{III}

To extract the global polarization from the previous analysis, a crucial ingredient is the calculation of the intrinsic polarizations $z$ and $\bar{z}$.  Following the analysis in Refs.~\cite{Ayala:2020soy,Ayala:2020ndx},  the intrinsic polarizations are given by 
\begin{eqnarray}
    z &=& 1 - e^{-\Delta \tau_{QGP}/\tau}\nonumber\\
    \bar{z} &=& 1 - e^{-\Delta \tau_{QGP}/\bar{\tau}},
    \label{eqz}
\end{eqnarray}
\begin{figure}
    \centering
    \includegraphics[width=0.5\textwidth]{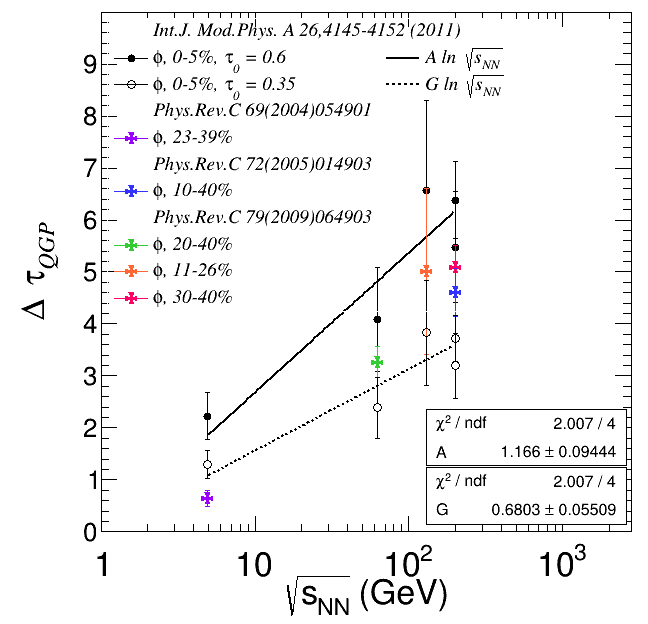}
    \caption{The QGP lifetime $\Delta \tau_{QGP}$ as a function of collision energy for central and $\approx$ 10\% - 40\% collisions. Empty points are calculated with $\tau_0 = 0.35$ fm and for the filled points with $\tau_0 = 0.6$ fm for $T_0$ extracted from $\phi$ spectra measured at central collisions 0\%-5\% ~\cite{Sahoo:2010gr}; the dashed and black line are the corresponding fits which delimit a region in which the $\Delta \tau_{QGP}$ estimated for non-central collisions are located (colored points). They are obtained with the corresponding $T(\tau_0)$ measured in different experiments~\cite{Back:2003rw,Abelev:2008aa,Adler:2004hv} and $\tau_0 = 0.5$ fm.}
    \label{fig:timeqgp}
\end{figure}
in terms of the relaxation times $\tau$ and $\bar{\tau}$ for the alignment between the spin of a quark $s$ or a $\bar{s}$ with the thermal vorticity, and within the QGP lifetime $\Delta \tau_{QGP}$. Equations~(\ref{eqz}) assume that the $s$ and $\bar{s}$ quark polarizations translate into the $\Lambda$ and $\overline{\Lambda}$ polarization, respectively, during the hadronization process. The relaxation times $\tau$ and $\bar{\tau}$ can be computed as the inverse of the interaction rate for the spin alignment of a massive quark or antiquark with energy $p_0$ with the angular velocity with magnitude $\omega$ as~\cite{Ayala:2020ndx}
\begin{eqnarray}
\Gamma(p_0)&=& \omega^2 \Gamma^\prime (p_0)
\end{eqnarray}
with
\begin{eqnarray}
    \Gamma^\prime (p_0) &=& \frac{\alpha_s}{4 \pi T^2} \frac{C_F}{\sqrt{p_0^2-m_q^2}}\int_0^\infty dk k \int_{\mathcal{R}} dk_0[1+f(k_0)] \nonumber \\
    && \tilde{f}(p_0+k_0-\mu_q)\sum_{i=L,T} C_{i}(p0,k_0,k)\rho_i(k_0,k). \nonumber \\
    \label{intrate}
\end{eqnarray}
where the integral is performed over the kinematical available region, weighted with the relevant statistical distributions the Bose-Einstein $f$ and the Fermi-Dirac, $\tilde{f}$ for gluons and quarks, respectively. $C_i$, $i=T,\ L$ are the result of the trace calculation after contraction of the transverse and longitudinal projection operators --that come together with the gluon spectral functions $\rho_i$-- with the quark propagator and the vertices, after summing over the Matsubara frequencies (see Ref.~\cite{Ayala:2020ndx} for further details).
\begin{figure}
    \centering
    \includegraphics[width=0.5\textwidth]{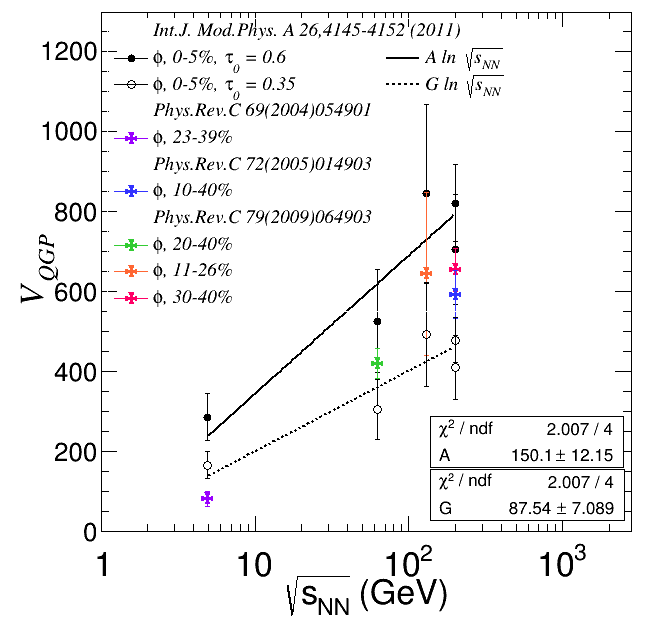}
    \caption{Volume of the QGP as a function of the collision energy for central and $\approx$ 10\% - 40\% collisions. The region delimited by the fits to the volume estimated from data at central collisions with $\tau_{0} = {0.60, 0.35}$ fm, corresponds to the volume calculated with data at different centralities $\approx$ 10\% - 40\% and $\tau_0 = 0.5$ fm as we can see indicated by the colored points.}
    \label{fig:Volume}
\end{figure}
The total interaction rate is obtained integrating Eq.~(\ref{intrate}) over the quark phase space and is given by  
\begin{eqnarray}
	\Gamma &=& V \omega^2 \int \frac{d^3p}{(2 \pi)^3}\Gamma^\prime(p_0),
\end{eqnarray}
where $V$ represents the volume of the core region. 

To compute $V$ and $\Delta\tau_{QGP}$ for conditions that depend on the collision energy, we consider a  Bjorken expansion scenario where the volume and the QGP life-time are related by
\begin{equation}
 V = \pi R^2 \Delta \tau_{QGP},   
\end{equation}
where $R$ is the radius of the colliding species. The QGP life-time is given as the interval elapsed from the initial formation $\tau_0$ until the hadronization time $\tau_f$. There is no unique way to estimate $\tau_0$ and $\tau_f$. For these purposes, both electromagnetic and hadron probes (data and simulation) have been used in the literature to provide complementary information to  estimate these times. In this work we assume and ideal fluid made out of quarks and gluons undergoing a Bjorken expansion~\cite{QGPlectures,QGPlectures2} and thus relating these times to the corresponding fluid temperatures $T_{f} = T(\tau_{f})$ and $T_0=T(\tau_0)$ by means of
\begin{equation}
    \Delta \tau_{QGP} = \tau_f - \tau_0 = \tau_0 \left[ \left(\frac{T_0}{T_f} \right)^3 -1 \right].
\end{equation}

$T_f$ is obtained from Eq.~(\ref{Tmurel}) for different values of $\mu_B$. To estimate $T_0$, we use data from the transverse momentum of $\phi$-mesons~\cite{Sahoo:2010gr}. We consider a range of values of $\tau_0=0.35 - 0.60$ fm to incorporate the effect of the collision centrality on the initialization of the QGP formation. This is a reasonable range of values for $\tau_0$ that is also consistent with the estimated initial temperature $T_0$~\cite{hydro,hydro2,hydro3}. Figures \ref{fig:timeqgp} and \ref{fig:Volume} show the QGP life-time and volume as a function of the collision energy for central collisions ($0 - 5 $)\% evaluated with $\tau_0=0.35$ fm and $\tau_0=0.60$ fm. This is equivalent to evaluate the life-time and volume of the QGP for other centralities, as we can see from the fits to these data, which delimit a region that contains the QGP life-time and volume estimated with $\tau_{0} = 0.5$ fm and $T_{0}$ extracted from $\phi$ mesons produced in collisions at $10$\% - $40$\% of centrality.

To estimate $\omega$ for the appropriate value of the impact parameter ($b = 7.26$ fm), we use a linear interpolation of the ones reported in Ref.~\cite{Deng:2020ygd,Deng:2016gyh}) for Au + Au collisions, as a function of $\sqrt{s_{NN}}$ and impact parameters $b = {5,\ 8,\ 10}$ fm.

\begin{figure}[t]
    \centering
    \includegraphics[width=0.45\textwidth]{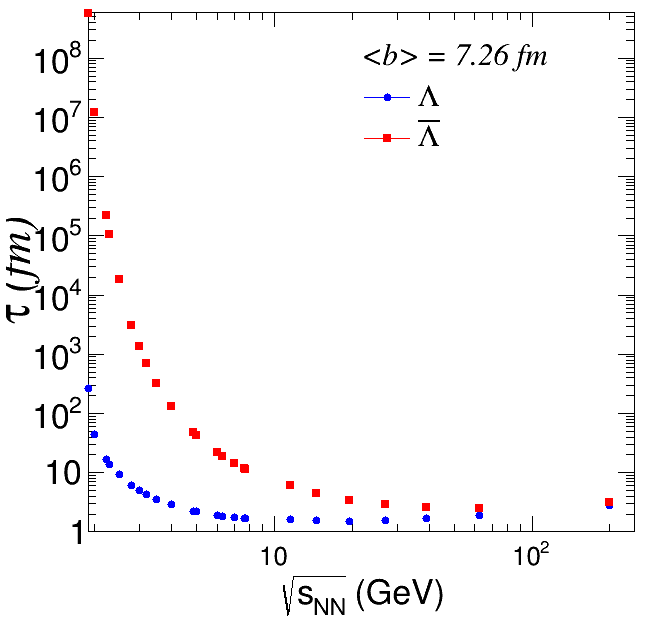}
    \caption{Relaxation times $\tau$ ($\bar{\tau}$) for $\Lambda$ ($\overline{\Lambda}$) Corresponding to the QGP volume evaluated with $\tau_{0} = 0.60$ fm.}
    \label{fig:time}
\end{figure}

\begin{figure}[t]
    \centering
    \includegraphics[width=0.45\textwidth]{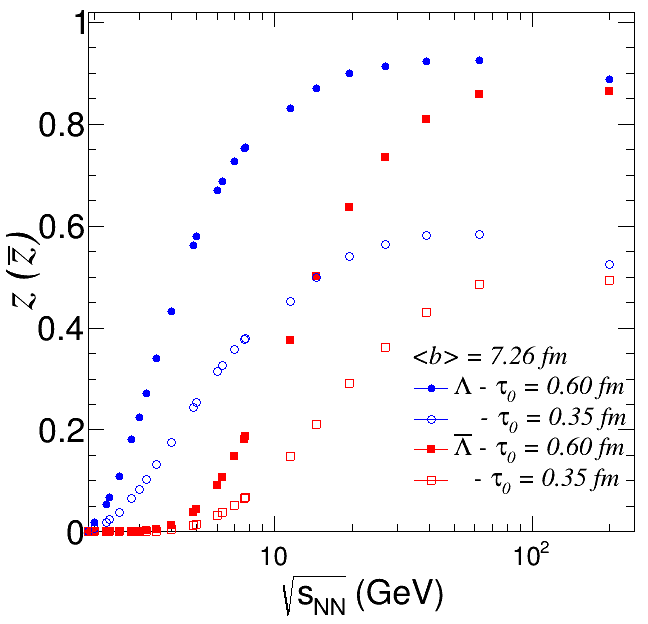}
    \caption{Distributions of $z$ for two different values of $\tau_{0}$ that corresponds to the centrality of HADES measurement for $\Lambda$ and $\bar{\Lambda}$ hyperon global polarization.}
    \label{fig:intrinsic}
\end{figure}

Using the total interaction rate $\Gamma$, the volume of the overlap region $V$, the QGP life-time $\Delta\tau_{QGP}$ and the angular velocity estimation of $\omega$, we can obtain the relaxation times as $\tau \equiv 1/\Gamma(\mu_B)$ and $\bar{\tau} \equiv 1/\Gamma(-\mu_B)$. Figure~\ref{fig:time} shows the relaxation times thus obtained. Notice that for energies below the $\Lambda$-production threshold energy, the relaxation times increase dramatically, as expected, since the interaction rate should vanish below these energies. We can now use Eq.~(\ref{eqz}) to calculate the intrinsic polarizations $z$ and $\bar{z}$. These are shown in Fig.~\ref{fig:intrinsic}. Notice that $z$ drops down to values close to zero for energies below $\sqrt{s_{NN}} \approx 5$ GeV.

\begin{figure}[t]
    \centering
    \includegraphics[width=0.45\textwidth]{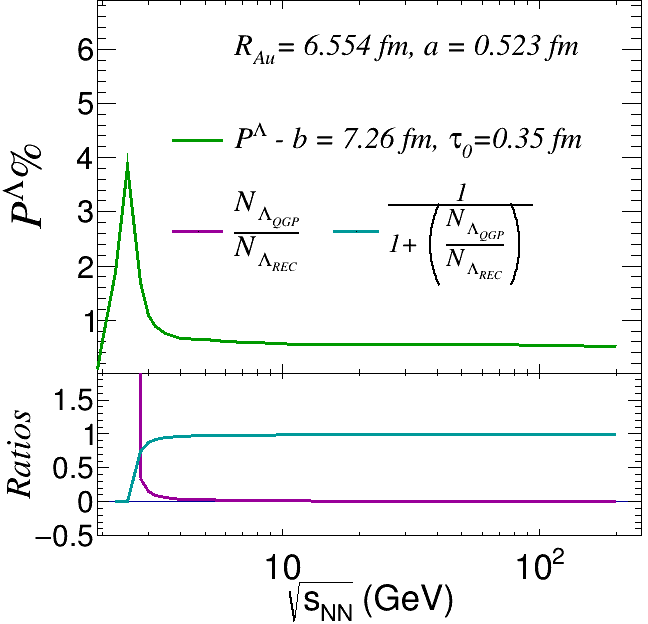}
    \includegraphics[width=0.45\textwidth]{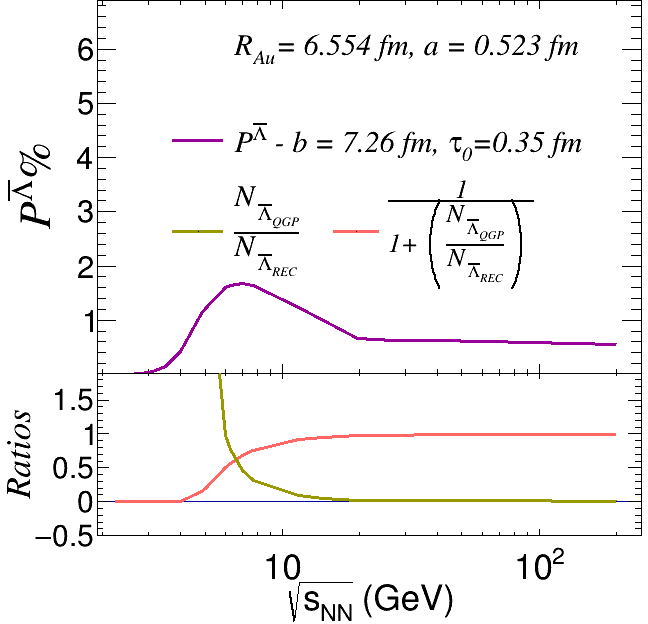}
    \caption{Global polarizations ${\mathcal{P}}^\Lambda$ (top panel) and ${\mathcal{P}}^{\overline{\Lambda}}$ (bottom panel) as functions of collision energy an for fixed values of the model parameters. Also shown are the monotonically decreasing ratio $N_{\Lambda\ {\mbox{\tiny{QGP}}}}/N_{\Lambda\ {\mbox{\tiny{REC}}}}$ ($N_{\overline{\Lambda}\ {\mbox{\tiny{QGP}}}}/N_{\overline{\Lambda}\ {\mbox{\tiny{REC}}}}$) and the monotonically increasing ratio $1/(1+N_{\Lambda\ {\mbox{\tiny{QGP}}}}/N_{\Lambda\ {\mbox{\tiny{REC}}}})$ ($1/(1+N_{\overline{\Lambda}\ {\mbox{\tiny{QGP}}}}/N_{\overline{\Lambda}\ {\mbox{\tiny{REC}}}})$), which are the parameters that provide the peaking behavior of the polarization functions. In fact, notice that the polarizations peak near where these ratios cross each other. The exact location of the peak is controlled by the energy dependence of $z$ ($\bar{z}$).}
    \label{fig:comparison}
\end{figure}

Before proceeding to a detailed study of the $\Lambda$ and $\overline{\Lambda}$ polarization excitation functions aimed to be compared to experimentally available data, we first show that by putting together these ideas, we can describe a main feature of these excitation functions, namely, the existence of peaks for both of them at given, albeit different, collision energies. Figure~\ref{fig:comparison} shows the global polarizations ${\mathcal{P}}^\Lambda$ (top panel) and ${\mathcal{P}}^{\overline{\Lambda}}$ (bottom panel) as functions of the collision energy, for fixed values of the model parameters. The figure also shows the behavior of the monotonically decreasing ratio $N_{\Lambda\ {\mbox{\tiny{QGP}}}}/N_{\Lambda\ {\mbox{\tiny{REC}}}}$ ($N_{\overline{\Lambda}\ {\mbox{\tiny{QGP}}}}/N_{\overline{\Lambda}\ {\mbox{\tiny{REC}}}}$) and the monotonically increasing ratio $1/(1+N_{\Lambda\ {\mbox{\tiny{QGP}}}}/N_{\Lambda\ {\mbox{\tiny{REC}}}})$ ($1/(1+N_{\overline{\Lambda}\ {\mbox{\tiny{QGP}}}}/N_{\overline{\Lambda}\ {\mbox{\tiny{REC}}}})$), which, according to Eq.~(\ref{eq1-mod}), are the ratios that provide the main energy behavior of the polarization functions. Notice that the global polarizations peak near where these functions cross each other. The position of the peaks are slightly displaced from these crossing points since the intrinsic polarizations $z$ and $\bar{z}$ also have a (mild) energy dependence.

\section{Excitation function for the global $\Lambda$ and $\overline{\Lambda}$ polarization}\label{IV}

We use the previous results to calculate the global $\Lambda$ and $\overline{\Lambda}$ polarization as functions of energy in centrality intervals that are relevant to the STAR-BES and the HADES measurements.

\begin{figure}
    \centering
    \includegraphics[width=0.45\textwidth]{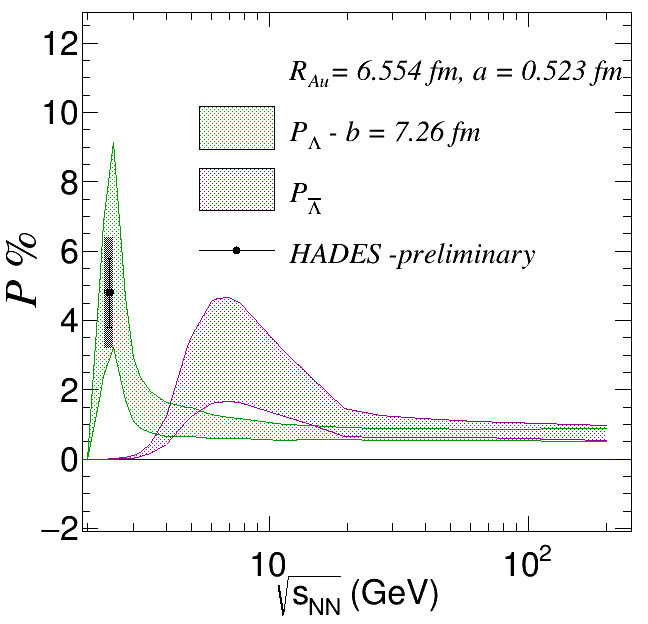}
    \includegraphics[width=0.45\textwidth]{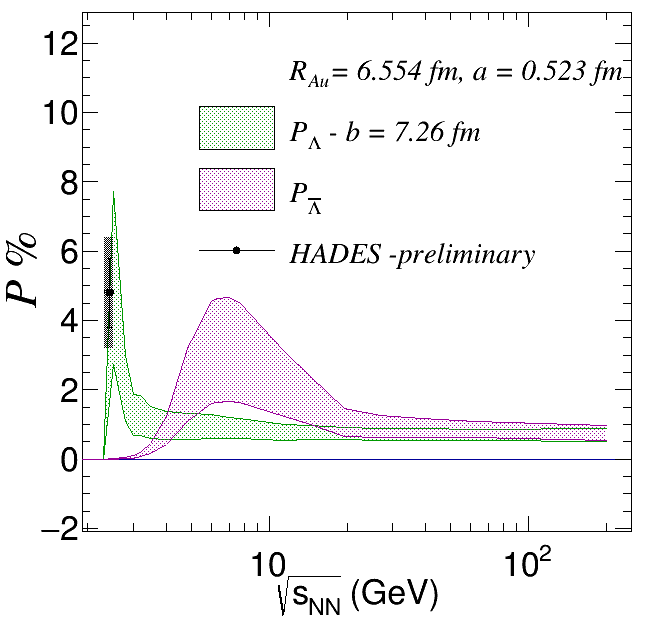}
    \caption{Polarization as a function of the collision energy. For $\sqrt{s_{NN}} < 5$ GeV we use different fits to the nucleon-nucleon inelastic cross-section $\sigma_{NN}$ and for higher energies we use the reported in~\cite{Nakamura:2010zzi}. Upper panel shows results with \textit{Fit 1}~\cite{Buss:2011mx} and lower panel shows results with \textit{Fit 2}~\cite{Kardan:2015,Bystricky:1987yq}. Both panels show  preliminary data point from HADES as reported in Ref.~\cite{SQM2019}. Shaded areas correspond to the region  delimited  by the values of $z$ and $\bar{z}$ calculated with  the  fits  to  the  QGP volume and lifetime as shown in Figs.~(\ref{fig:timeqgp}) and (\ref{fig:Volume}).}
    \label{fig:polvsen}
\end{figure}

Figure~\ref{fig:polvsen} shows the polarization computed for $b = 7.26$ fm corresponding to the centrality range 10\% - 40\% which is the range used for the HADES preliminary measurement~\cite{SQM2019}. For $\sqrt{s_{NN}} \leq 7.0$ GeV 
we use two different fits for $\sigma_{NN}$. The result for the \textit{Fit 1}~\cite{Buss:2011mx} is shown in the upper panel and for the \textit{Fit 2}~\cite{Bystricky:1987yq} in the lower panel. For higher energies, we use the parametrization reported in Ref.~\cite{Nakamura:2010zzi}, according to the discussion in Sec.~\ref{II}. The behaviour of the polarization excitation functions is similar and the difference is more noticeable for the height of the $\Lambda$ polarization at small collision energies. 
For higher energies, the trend is in agreement with the STAR-BES results. Notice that the  $\overline{\Lambda}$ polarization maximum is close to $\sqrt{s_{NN}}= $ 7.7 GeV. On the other hand, for HADES energies and the centrality range 10\% - 40\%, the $\Lambda$ polarization maximum is close to $\sqrt{s_{NN}} \approx 2.5$ GeV. This energy corresponds to the threshold energy for $\Lambda$ production in the $p + p$ $\rightarrow \Lambda + K^-$ + $p$ channel. The results for the $\Lambda$ and  $\overline{\Lambda}$ polarizations are very similar for the two fits that we used for $\sigma_{NN}$.

\begin{figure}[t]
    \centering
    \includegraphics[width=0.45\textwidth]{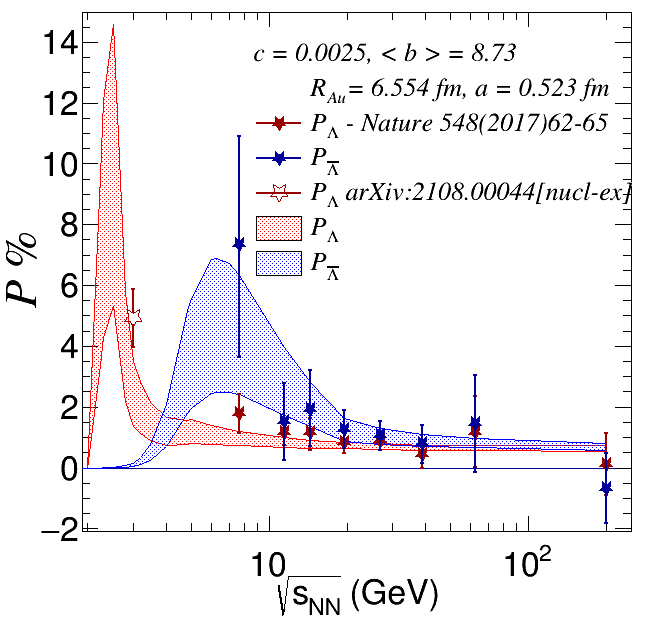}
    \caption{Polarization as a function of the collision energy for centrality range 20\% - 50\%. Comparison with STAR-BES data~\cite{STAR-Nature,STAR:2021beb}. Shaded areas correspond to the region  delimited  by  the  fits  to  the  QGP volume and lifetime as shown in Figs.~(\ref{fig:timeqgp}) and (\ref{fig:Volume}). }
    \label{fig:polbes}
\end{figure}

For the STAR-BES centrality range 20\% - 50\%, the average impact parameter is around the value at which the critical density is achieved and, consequently, the number of $\Lambda$s from the core changes drastically with small variations on either $b$ or $\sigma_{NN}$. Therefore, instead of using a single value for $b$, we compute the number of $\Lambda$s for a finite centrality range.

This range is computed using the geometric relation between the impact parameter and centrality given by~\cite{Broniowski:2001ei}:
\begin{equation}
    c(b) = \frac{\pi b^2}{\sigma_{AuAu}}\times 100 \%,
\end{equation}
where $\sigma_{AuAu}$ is the inelastic cross-section of the collision. Therefore 
\begin{eqnarray}
\langle b \rangle &=& \frac{1}{c_{f}-c_{i}}\int_{c_{i}}^{c_{f}} b(c) dc,
\label{avgb}
\end{eqnarray}
which yields $b_{20\%} \approx 6.66$ fm, $b_{50\%} \approx 10.52$fm, and $\langle b \rangle = 8.73$ fm. Thus, the average number of $\Lambda$s produced in the QGP and the corona, $\langle \NQGP \rangle$ and $\langle \NREC \rangle$  are given by
\begin{eqnarray}
    \langle \NQGP \rangle &=& \frac{1}{b_{50\%} - b_{20\%}}\int_{b_{20\%}}^{b_{50\%}} \NQGP(b)db, \nonumber \\
    \langle \NREC \rangle &=& \frac{1}{b_{50\%} - b_{20\%}}\int_{b_{20\%}}^{b_{50\%}} \NREC(b)db.
\end{eqnarray}
Using these results into Eq.~(\ref{eq1}) and calculating the intrinsic polarization with the mean value of the impact parameter $\langle b \rangle$ in Eq.~(\ref{avgb}), we obtain the polarization for the STAR-BES centrality range. This is shown in Fig.~\ref{fig:polbes}. Notice that our analysis provides an excellent description STAR-BES data~\cite{STAR-Nature} over the entire collision energy range, including also the latest polarization value at $\sqrt{s_{NN}}=3$ GeV~\cite{STAR:2021beb}, reported after our study was first released. We observe that the trend is similar to the case of the analysis with a smaller centrality range. The difference is in the magnitude of the global polarization, which increases for larger centrality, as a consequence of the angular velocity increase.

In both Figs.~(\ref{fig:polvsen}) and~(\ref{fig:polbes}), the shaded areas correspond to the region  delimited  by  the  fits  to  the  QGP volume and life-time shown in Figs.~(\ref{fig:timeqgp}) and (\ref{fig:Volume}). Notice that in our approach, the space-time evolution of the QGP plays a central role in determining the height for the $\Lambda/\bar{\Lambda}$ polarizations.

\section{Summary and Conclusions}\label{V}

We have shown that the main characteristic features of the $\Lambda$ ($\overline{\Lambda}$) polarization excitation functions in semi-central relativistic heavy-ion collisions can be well described using a model where these hyperons come from a low density corona and a high density core regions, whose size and life-time depend on the collision energy. The main ingredient is shown to be the behavior of the product of the monotonically decreasing ratio $N_{\Lambda\ {\mbox{\tiny{QGP}}}}/N_{\Lambda\ {\mbox{\tiny{REC}}}}$ ($N_{\overline{\Lambda}\ {\mbox{\tiny{QGP}}}}/N_{\overline{\Lambda}\ {\mbox{\tiny{REC}}}}$) and the monotonically increasing ratio $1/(1+N_{\Lambda\ {\mbox{\tiny{QGP}}}}/N_{\Lambda\ {\mbox{\tiny{REC}}}})$ ($1/(1+N_{\overline{\Lambda}\ {\mbox{\tiny{QGP}}}}/N_{\overline{\Lambda}\ {\mbox{\tiny{REC}}}})$), which provide the prime energy behavior of the polarization functions. The global polarizations peak near where the functions cross each other. Notice that in the $\overline{\Lambda}$ case, the above ratios are driven by the energy-dependent parameters $w$ and $w'$, namely, on the ratios of the number of produced $\overline{\Lambda}$s and $\Lambda$s in the corona, and core regions, respectively. In particular since $w$ is defined only for energies larger than the threshold energy for $\overline{\Lambda}$ production in $p+p$ collisions, this threshold produces a shift of the energy at which the ratios $N_{\overline{\Lambda}\ {\mbox{\tiny{QGP}}}}/N_{\overline{\Lambda}\ {\mbox{\tiny{REC}}}}$ and $1/(1+N_{\overline{\Lambda}\ {\mbox{\tiny{QGP}}}}/N_{\overline{\Lambda}\ {\mbox{\tiny{REC}}}})$ cross each other, compared to the $\Lambda$ case. This effect makes the $\overline{\Lambda}$ polarization peak at a larger energy than the $\Lambda$ polarization.

The other important ingredient that provides, in particular, the precise position of the peaks, is the relaxation time from which the intrinsic polarizations are computed. We have shown that these can be obtained from a field theoretical approach that links the alignment of the strange quark spin with the thermal vorticity, modeling the QGP volume and life-time using a simple scenario. Thus, the main finding of this work is the prediction of a maximum for the $\Lambda$ and $\overline{\Lambda}$ polarizations which should be possible to be measured in the NICA and HADES energy range.

It is worth emphasizing that, in our improved core-corona model, the scenario we put forward for the QGP production and its evolution (volume and life-time), are not the only two features to account for when applying the model to hyperon production. A key ingredient, the ratio $N_{\Lambda\ {\mbox{\tiny{QGP}}}}/N_{\Lambda\ {\mbox{\tiny{REC}}}}$, turns out to be highly sensitive on the centrality ranges, which in turn are defined in terms of the participants of the collision after using a Glauber model with associated impact parameter ranges. This means in particular that $\Lambda$s and $\overline{\Lambda}$s can still be produced, even if the mean impact parameter $<b>$ [see Eq.~(\ref{avgb})] is above the critical value ($b_c \approx 7.26$) to produce the QGP. Furthermore, we know that the volume of the QGP increases with collision energy, as shown in Fig.~\ref{fig:Volume}. However, the number of $\Lambda$s produced in the core ($N_{\Lambda\ {\mbox{\tiny{QGP}}}}$) do not follow this trend. In fact, $N_{\Lambda\ {\mbox{\tiny{QGP}}}}$ grows quadratically with the number of participants in the collision ($\NpQGP$), as shown by Eqs.~(\ref{numberof participants}) and~(\ref{NLQGP}), whereas $\NpQGP$ shows a steady but small growth beyond NICA energies coming from the collision energy-dependent nucleon + nucleon cross-section $\sigma_{NN}$. On the other hand, $\Lambda$ production in the corona ($N_{\Lambda\ {\mbox{\tiny{REC}}}}$) is proportional to the  nucleon + nucleon cross-section $\sigma_{pp}^{\Lambda}$ [see Eq.~(\ref{partper})] for which the fit to data is described in terms of a logarithmic growth, as shown in Fig.~\ref{fig:sigmafit}. In broad terms, this provides a differential lambda production growth with collision energy: in the core it tends to stabilize, whereas in the corona it tends to grow with energy, for different impact parameter ranges. 

Recently, the RHIC-BES analysis on $\Lambda$ yields at different centralities~\cite{STAR:2019bjj}, show that there is a decrease in $\Lambda$ production for central collisions ($5\%-10\%$) when going from 7.7 GeV up to 39 GeV in collision energy. This behavior is different from the corresponding result on semi-peripheral and peripheral collisions ($40\%-60\%$ and above), which show no apparent energy dependence. The explanation for this behavior, also mentioned in Ref.~\cite{STAR:2019bjj}, may be linked to an increase of baryon density in the collision system, which in turn comes from an increase in baryon stopping. Altogether, these results call for further analysis to improve the scenario of hyperon production in the QGP. We are currently pursing these studies and we will report our findings elsewhere.

\section{Acknowledgements} 

I.M. thanks the ICN-UNAM faculty and staff for the support and kind hospitality provided during the development of part of this work and acknowledges support from a postdoctoral fellowship granted by CONACyT, M\'exico. Support for this work has been received by UNAM-DGAPA-PAPIIT grant number IG100322 and by CONACyT grant numbers A1-S-7655 and A1-S-16215.

\bibliography{Bibliografia}

\end{document}